# Interparticle heterostructures by spontaneous formation of Dirac nodal arc semimetal $PtSn_4$ domains in nanoparticles produced by Supersonic Cluster Beam Deposition.

Marc Heggen[1], José Enrique Martinez Medina[2,3], Emanuele Barborini[2,*]

[1]Ernst Ruska-Centre for Microscopy and Spectroscopy with Electrons (ER-C), Forschungszentrum Jülich GmbH, Jülich, Germany
[2]Luxembourg Institute of Science and Technology (LIST), Belvaux, Luxembourg
[3]Department of Physics and Materials Science, University of Luxembourg, Esch-Sur-Alzette, Luxembourg

In this study, we report on the spontaneous formation of highly ordered 2D layered domains of intermetallic phase $PtSn_4$ in Sn nanoparticles of dimensions of the order of 10 nm during the gas aggregation process occurring in Supersonic Cluster Beam Deposition. Phase identification is based on High Resolution Transmission Electron Microscopy and on X-ray emission analysis coupled with Scanning Transmission Electron Microscopy. We propose that PtSn4-ordered domains precipitate inside Sn nanoparticles once the temperature drops below 520°C upon collisional cooling with room temperature Argon, while the nanoparticles persist longer in a liquid state. Sn matrix eventually solidifies upon the sudden temperature drop due to the supersonic expansion. 2D-layered $PtSn_4$ domains create interparticle heterostructures that disrupt the spherical symmetry typical of gas aggregation processes and separate the Sn particle into distinct parts. The Dirac nodal arc semimetal character of $PtSn_4$ makes it particularly interesting for studying the transport mechanisms in nanogranular films obtained by the soft-assembling of such nanoparticles, which feature a network of heterostructures showing sequences of alternate $PtSn_4$ 2D domains and metallic β-Sn necks.

Platinum and tin exhibit a range of ordered intermetallic phases, including $Pt_3Sn$, $Pt_2Sn_3$, $PtSn$, $PtSn_2$, and $PtSn_4$. Among them, $PtSn_4$ stands out as a phase of great interest from both a fundamental and applicative point of view, because of its topological semimetal character with a 2D crystal structure and associated unique properties [1,2]. The presence of Dirac-like electronic states and nodal arc features in its band structure makes $PtSn_4$ of interest for spintronics and might explain the extreme magnetoresistance observed, which makes it a fascinating material for research on topological physics and electronic transport phenomena [3,4]. Additionally, the interplay between its layered structure and electron transport behavior has implications for thermoelectric applications, where low thermal conductivity combined with high electrical conductivity is desirable [5].

The layered arrangement of the Pt and Sn atoms of $PtSn_4$ provides surface active sites with modified electronic states, suggesting catalytic potential in electrochemical reactions such as the hydrogen evolution reaction (HER) and the oxidation processes in fuel cells. In this regard, Li et al. propose that $PtSn_4$ might at the same time have surface state robustness with respect to chemical reactions and good carrier density and mobility typical of Dirac nodal arc semimetals. The resulting high electrical conductivity provides fast charge transport and decreases the Schottky barrier at the catalyst-adsorbate interface. Combined with the reduced amount of platinum, these properties feature $PtSn_4$, a sustainable catalyst of remarkable interest [6,7].

Beynon et al. recently reported for the first time on the synthesis of $PtSn_4$ in the form of a crystalline thin film by electron beam evaporation and provided its full characterization, emphasizing the importance of their achievement for quantum technologies [8]. In the framework of low dimensionality systems, nanostructuring would provide advantages in terms of reactive area and insights into the evolution of the Dirac nodal arc semimetal character with the dimension of the crystalline domains, however, the synthesis and stabilization of $PtSn_4$ at the nanoscale remain challenging, and studies on its nanostructured forms are still limited. One of the major breakthroughs was reported by Yu et al., who recently described the production of heterostructured nanoparticles by a wet-chemistry approach based on the galvanic replacement method, where various segregated intermetallic phases with 2D symmetry are created from tin nanoparticles [9]. Besides the description of the synthesis method, they highlight that layered structures of intermetallic phases break the spherical symmetry of tin nanoparticles and create interparticle platelet-like heterostructures.

[*]Corresponding author: Emanuele Barborini, Luxembourg Institute of Science and Technology (LIST), 41 Rue du Brill, L-4422 Belvaux, Luxembourg, emanuele.barborini@list.lu

In contrast to wet chemistry, the gas-phase synthesis of metallic nanoparticles is a conceptually simple approach featuring high control of the material composition and contamination, which proceeds from the vaporization of a solid-state target and the subsequent nucleation of particles and growth by the isotropic addition of monomers. The absence of any other chemical compound, apart from the atomic precursors and the high-vacuum environment in which the process takes place, results in excellent control of the composition and surface termination of the nanoparticles. When applied to multi-component materials, the gas-phase synthesis of nanoparticles evolves from the initial state characterized by the uniform distribution of the atomic precursors in the starting vapor, to the final condensed state where the different precursors may become segregated within the nanoparticles. This gives rise to core-shell or Janus structures, for instance, or to other complex configurations, depending, to a largely unknown extent, on the nature of the atomic species undergoing aggregation, as well as the way in which the initial vapor is quenched to induce nucleation, growth, and liquid-solid phase transition [10,11].

Aiming to study the gas-phase synthesis of intermetallic Sn-Pt phases in the form of ultrasmall nanoparticles, we adopted Supersonic Cluster Beam Deposition (SCBD) equipped with Pulsed Microplasma Cluster Source (PMCS) [12-14]. The method implements a gas aggregation process by synchronizing a pulsed electrical discharge with a pulsed jet of inert gas that acts first as a sputtering promoter of the metallic precursor target, in the form of a localized plasma plume, then as a thermal bath for metallic atoms to cool down and aggregate in nanoparticles, and finally extracting the nanoparticles from the source through a nozzle supersonic gas expansion towards a vacuum chamber. The Isentropic character of supersonic expansion causes the temperature of the inert gas to suddenly drop downstream the nozzle, freezing the crystalline structures of carried nanoparticles. PMCS was fed with argon as a process gas and equipped with a target made of sintered tin and platinum with an Sn-Pt composition of 85-15 at%. Under the simplistic hypothesis of a similar sputtering yield of Pt and Sn, this composition provides a vapor in the region of the Sn-Pt phase diagram where $PtSn_4$ is the thermodynamically favored condensed phase [15]. However, since the sputtering yield of Sn exceeds that of Pt, an excess of Sn is expected in the final nanoparticle composition. We collected the nanoparticles by directly exposing standard transmission electron microscopy (TEM) grids to the supersonic cluster beam. The deposition system operated at a base pressure of about $5 \times 10^{-8}$ mbar.

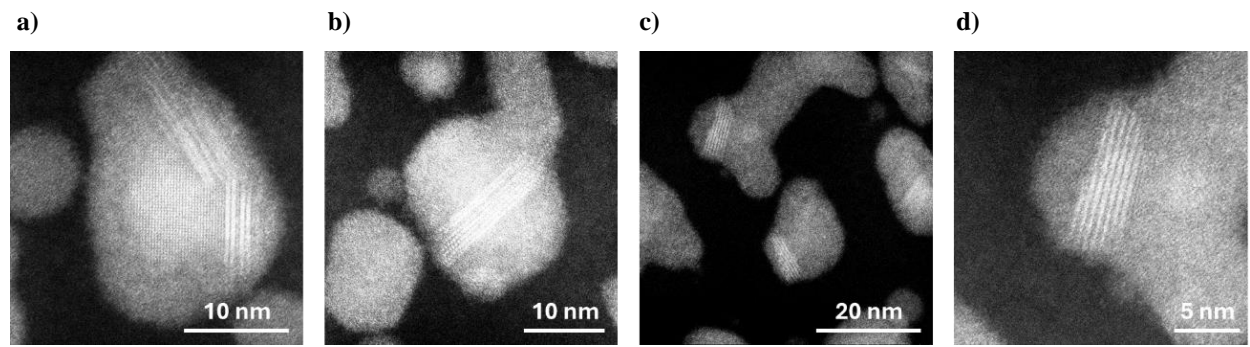

FIG 1. Various HAADF-STEM images of nanoparticles produced by SCBD starting from a solid precursor target made of sintered tin-platinum at 85% and 15% atomic composition, respectively. Interparticle-layered crystalline domains with bright contrast, breaking the spheroidal symmetry of the nanoparticles, are clearly visible. Image d) is an enlarged image of the top-left object of image c).

Scanning transmission electron microscopy (STEM) was conducted on a Hitachi HF5000 equipped with a spherical-aberration probe corrector operating at 200 kV. In high-angle annular-dark-field (HAADF) mode, STEM images unveil bright contrast layered domains, appearing as clearly segregated elements embedded into nanoparticles, breaking their spheroidal symmetry (Fig. 1). The presence of particles with cores characterized by a bright contrast (although less intense than that of layered structures) and evidence of partial coalescence among nanoparticles are also observed. As the STEM analysis is conducted ex-situ, core-shell structures visible in the larger particles originate from the Cabrera-Mott oxidation, leaving the unreacted metallic β-Sn phase in the core surrounded by an SnOx shell. The presence of core-shell structures in the particles with a dimension beyond a certain threshold, the observation of a uniform contrast in smaller particles matching that of shells and suggesting full oxidation, and the evidence of partial coalescence occurring among nanoparticles are in line with our recent results on the structure of SnOx nanoparticles by SCBD and of the nanogranular films obtained by Sn nanoparticle soft assembling [16-18].

Elemental analysis by EDX-STEM was conducted with an Aztec EDX system using two Ultim Max Silicon Drift Detectors (Oxford Instruments, Abingdon, UK). Although quantitative conclusions cannot be drawn, the elemental composition of the nanoparticles clearly shows that the quantity of Sn with respect to Pt exceeds the one of the sintered target, as expected from the different sputtering rate. This suggests that the composition of the initial gas-phase state is positioned in the region of the Sn-Pt phase diagram, where the coexistence of the metallic β-Sn phase and intermetallic $PtSn_4$ phase is the thermodynamically favored condensed configuration. In line with this, the inter-particle distribution of Sn and Pt shows a remarkable segregation, where the former is uniformly distributed while the latter accumulates in correspondence with the layered domains with a bright contrast (Fig. 2).

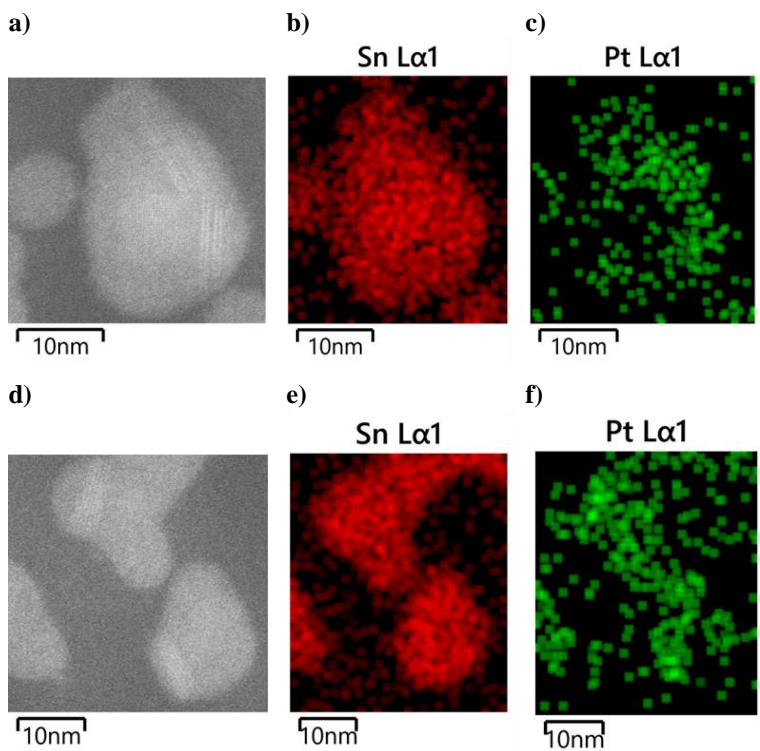

FIG 2. a-c) EDX-STEM analysis of the particle of Fig. 1(a): HAADF-STEM image (a), tin distribution based on Sn-Lα1 (b), and platinum distribution based on Pt-Lα1 (c). d-f) EDX-STEM analysis of the particles of Fig. 1(c). In all cases, while tin appears to be distributed uniformly within the nanoparticles, platinum tends to segregate, and its distribution is correlated with the brightly contrasting layered structures.

Fig. 3 shows the high-resolution STEM images of the particles shown in Fig. 1a-b. We identify a (almost) cubic pattern in the center of the nanoparticle with 2.8 and 2.9 Å lattice spacing and distinct, brightly contrasting layered domains with a 5.8 and 4.4 Å lattice parameter. The analysis of the high-resolution images unveils the presence of the $PtSn_4$ intermetallic phase oriented (101) with respect to the electron beam and β-Sn metallic phase oriented (001) surrounded by an amorphous, presumably tin oxide phase. Observations agree with the crystallographic parameters of the intermetallic $PtSn_4$ phase, which crystallizes in the orthorhombic space group Ccce and exhibits a layered structure composed of $PtSn_4$ sheets oriented along the (010) direction. Each Pt atom is 8-coordinate, bonded to eight equivalent Sn atoms in a distorted square antiprism geometry, with four shorter

Pt–Sn bonds (2.80 Å) and four slightly longer ones (2.81 Å).

Based on STEM and EDX-STEM results, we hypothesize that Sn and Pt atoms are released from the target upon an argon ion-pulsed sputtering aggregate in the form of small clusters, which grow by monomer addition and by coalescence, until they reach dimensions of the order of 10 nm. The difference in sputtering yields of Sn and Pt lead to a nanoparticle composition with a large excess of Sn. Given the low melting point of bulk tin (232°C) and considering the further decrease to values much lower than the bulk ones for particles with a diameter of about 10 nm [19], we propose that nanoparticles remain in the liquid state for long enough to allow the interparticle precipitation of the $PtSn_4$ crystalline phase (Fig. 4 a-b), which is predicted to exist in the solid state below 520°C [15]. A temperature above this value can therefore be inferred for the initial gas-phase state of Sn and Pt atomic precursors, upstream the aggregation process.

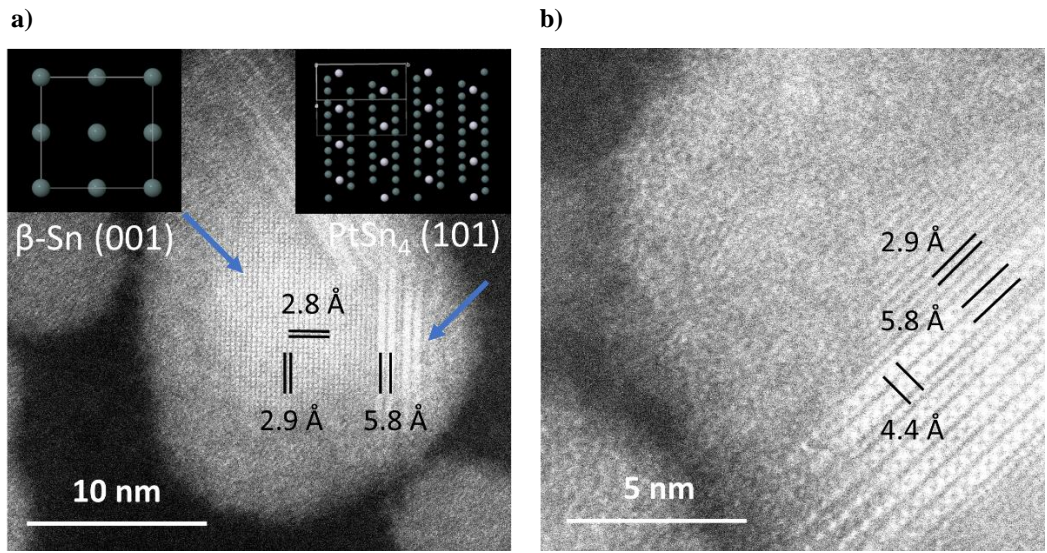

FIG 3. a) High-resolution HAADF-STEM image analysis of the particle of Fig. 1(a). A cubic pattern is observed in the center of the particle and layered domains on the side, the former with 2.8 and 2.9 Å lattice spacing and the latter with 5.8 and 4.4 Å. The cubic pattern is attributed to β-Sn metallic phase oriented (001) with respect to the electron beam and the layered pattern to the intermetallic phase $PtSn_4$ oriented (101). b) High-resolution HAADF-STEM image analysis of the particle of Fig. 1(b), leading to the same conclusions.

Sn-Pt particle cooling is initially provided by the collisions with argon within the PMCS body and subsequently during the supersonic expansion, for a total process duration—from target sputtering to supersonic beam formation—of the order of a few hundred microseconds. Although SCBD is sometimes considered an "out of equilibrium" process due to the sudden temperature drop of the supersonic expansion occurring immediately after particle formation and causing the freezing of their structures (and, for this reason, its potential ability to produce exotic structures is interesting), here we observe that the formation of highly ordered $PtSn_4$ domains has a dynamic fast enough to be compatible with the SCBD process timeframe and characteristics.

Interestingly, the 2D layered nature of the $PtSn_4$ phase not only creates heterostructures that disrupt the spherical symmetry of nanoparticles typically formed in gas aggregation processes, but in many observed instances, rather than being embedded within tin, the $PtSn_4$ phase sharply separates the Sn particles into two distinct parts. As we previously reported for cluster-assembled nanogranular Sn films deposited at room temperature [16], a similar partial coalescence is observed here as deposition progresses and substrate coverage increases (Fig. 4(c)). This can be linked to the dissipation of the kinetic energy of the particles within the supersonic argon beam, in combination with the reduced melting point of the material. As the deposition proceeds, this partial coalescence leads to the formation of an interconnected nanogranular network, with $PtSn_4$ layered domains segmenting the β-Sn metallic phase, as shown schematically in Fig. 4(d).

Given that this configuration incorporates Dirac nodal arc semimetal units that interrupt the nanogranular metallic network, we believe that the electrical transport properties of this system warrant specific in-situ investigations during the growth in high-vacuum conditions, to minimize the impact of oxidation. Phenomena such as quantum tunneling between non-connected particles at low coverage, the onset and evolution of percolation with increasing coverage, and resistive switching at high coverage may all be affected. Additionally, in-situ studies on the material's electronic band structure—examining the relationship between the Dirac semimetal characteristics of PtSn$_4$ and its nanostructured form—could benefit from the possibility of direct interfacing of the deposition system with surface science facilities thanks to the highly collimated nanoparticle beam and the room temperature deposition process associated with SCBD.

## ACKNOWLEDGMENTS
This work has been supported by the Luxembourg National Research Fund (FNR), CLASMARTS project (C19/MS/13685974), and by SNOX project (ER-C E-002). We acknowledge the support of Hitachi High Technologies.

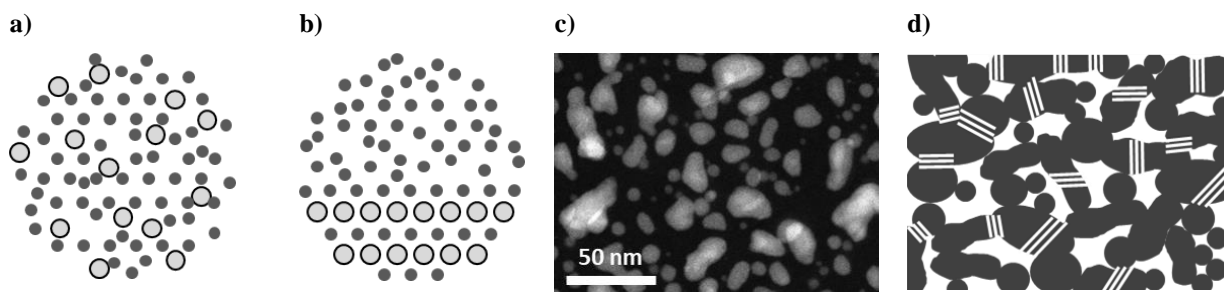

FIG 4. Schematic representation of Sn-Pt nanoparticles during the cooling process proceeding with gas-phase aggregation, and after deposition. a) Pt and Sn atoms (light and dark gray dots, respectively) are uniformly distributed in a nanoparticle in the liquid state, where Sn represents the largest component by far. According to the Sn-Pt phase diagram, the temperature here is expected to be higher than 520°C. b) Below 520°C, the ordered PtSn$_4$ phase precipitates within the still-liquid Sn nanoparticle, which eventually solidifies upon supersonic expansion cooling. c) During deposition, the partial coalescence of the nanoparticles occurs, bridging nanoparticles through the partial melting of β-Sn and forming a metallic tin nanogranular network (at a stage before percolation in the TEM image). d) Scheme of the possible final structure of the nanoparticle-assembled film where the β-Sn metallic nanogranular network (dark gray) is sectioned by PtSn$_4$ layered domains (white).

[1] Wu, Y., Wang, L.-L., Mun, E., Johnson, D. D., Mou, D. X., Huang, L., ... & Kaminski, A., Dirac node arcs in PtSn$_4$, Nature Physics, 12(7), 667-671 (2016).
https://doi.org/10.1038/nphys3712

[2] N. Kumar, S.N. Guin, K. Manna, C. Shekhar, C. Felser. Topological Quantum Materials from the Viewpoint of Chemistry. Chem. Rev. 2021, 121, 5, 2780–2815.
https://doi.org/10.1021/acs.chemrev.0c00732

[3] X. Luo, R. C. Xiao, F. C. Chen, J. Yan, Q. L. Pei, Y. Sun, W. J. Lu1, P. Tong, Z. G. Sheng, X. B. Zhu, W. H. Song, and Y. P. Sun, Origin of the extremely large magnetoresistance in topological semimetal PtSn4, Phys. Rev. B 97, 205132 (2018).
https://doi.org/10.1103/PhysRevB.97.205132

[4] E. Mun, H. Ko, G.J. Miller, G.D. Samolyuk, S.L. Bud'ko, and P.C. Canfield, Magnetic field effects on transport properties of PtSn4. Phys. Rev. B 85, 035135 (2012).
https://doi.org/10.1103/PhysRevB.85.035135

[5] Fu, C., Sun, J., Liu, Y., Jing, Y., Xu, X., Chen, X., ... & Zhao, X., Largely suppressed magneto-thermal conductivity and enhanced thermoelectric performance in PtSn$_4$, Nature Communications, 11(1), 1-8 (2020).
https://doi.org/10.34133/2020/4643507

[6] G. Li, C. Fu, W. Shi, L. Jiao, J. Wu, Q. Yang, R. Saha, M. E. Kamminga, A. K. Srivastava, E. Liu, A. N. Yazdani, N. Kumar, J. Zhang, G. R. Blake, X. Liu, M. Fahlman, S. Wirth, G. Auffermann, J. Gooth, S. Parkin, V. Madhavan, X. Feng, Y. Sun, C. Felser, Angew. Chem. Int. Ed. 58, 13107 (2019).
https://doi.org/10.1002/anie.201906109

[7] D.W. Boukhvalov, A. Marchionni, J. Filippi, C-N. Kuo, J. Fujii, R. Edla, S. Nappini, G. D'Olimpio, L. Ottaviano, C.S. Lue, P. Torelli, F. Vizza and A. Politano. Efficient hydrogen evolution reaction with platinum stannide PtSn4

[7] via surface oxidation. J. Mater. Chem. A 8, 2349–2355 (2020). https://doi.org/10.1039/C9TA10097K

[8] Beynon EL, Barker OJ, Veal TD, O'Brien L, O'Sullivan M., Heterostructure growth, electrical transport and electronic structure of crystalline Dirac nodal arc semimetal PtSn4, Scientific Reports 28;14(1):30887 (2024). https://doi.org/10.1038/s41598-024-81679-2

[9] Yu, J., Yin, Y. & Huang, W. Engineered interfaces for heterostructured intermetallic nanomaterials. Nat. Synth 2, 749–756 (2023). https://doi.org/10.1038/s44160-023-00289-4

[10] R. López-Martín, B.S. Burgos, P.S. Normile, J.A. De Toro, C. Binns. Gas Phase Synthesis of Multi-Element Nanoparticles. Nanomaterials 11, 2803 (2021). https://doi.org/10.3390/nano11112803

[11] J-G. Mattei, P. Grammatikopoulos, J. Zhao, V. Singh, J. Vernieres, S. Steinhauer, A. Porkovich, E. Danielson, K. Nordlund, F. Djurabekova, M. Sowwan. Gas-Phase Synthesis of Trimetallic Nanoparticles. Chem. Mater. 31, 6, 2151–2163 (2019). https://doi.org/10.1021/acs.chemmater.9b00129

[12] E. Barborini, P. Piseri and P. Milani, A pulsed microplasma source of high intensity supersonic carbon cluster beams, Journal of Physics D: Applied Physics 32, L105 (1999). https://doi.org/10.1088/0022-3727/32/21/102

[13] P. Milani, S. Iannotta, Cluster Beam Synthesis of Nanostructured Materials, Springer-Verlag, Berlin Heidelberg, 1999. https://doi.org/10.1007/978-3-642-59899-9

[14] E. Barborini, S. Vinati, Supersonic Cluster Beam Deposition for the integration of functional nanostructured films in devices, in: S. Krishnamoorthy, K. Iniewski (Eds.), Advances in Fabrication and Investigation of Nanomaterials for Industrial Applications, Springer-Nature Switzerland AG (2024). https://doi.org/10.1007/978-3-031-42700-8_1

[15] Cahn, Robert Wolfgang. Binary Alloy Phase Diagrams–Second edition. T. B. Massalski, Editor-in-Chief; H. Okamoto, P. R. Subramanian, L. Kacprzak, Editors. ASM International, Materials Park, Ohio, USA (1990). Advanced Materials 3 (1991): 628-629. https://doi.org/10.1002/adma.19910031215

[16] J.E. Martinez Medina, J. Polesel-Maris, A.M. Philippe, P. Grysan, N. Bousri, S. Girod and E. Barborini, The role of coalescence and ballistic growth on in-situ electrical conduction of cluster-assembled nanostructured Sn films. Applied Surface Science 664, 160268 (2024). https://doi.org/10.1016/j.apsusc.2024.160268

[17] J.E. Martinez Medina, A.M. Philippe, J. Guillot, C. Vergne, Y. Fleming and E. Barborini. Intermediate tin oxide in stable core-shell structures by room temperature oxidation of cluster-assembled nanostructured Sn films, Applied Surface Science 658, 159846 (2024). https://doi.org/10.1016/j.apsusc.2024.159846

[18] M. Heggen, J.E. Martinez Medina, A.M. Philippe and E. Barborini, Experimental insights on the stability of core–shell structure in single Sn/SnOx spherical nanoparticles during room temperature oxidation, Applied Surface Science 684, 161984 (2025). https://doi.org/10.1016/j.apsusc.2024.161984

[19] S.L. Lai, J.Y. Guo, V. Petrova, G. Ramanath and L.H. Allen, Size-Dependent Melting Properties of Small Tin Particles: Nanocalorimetric Measurements, Physical Review Letters 77, 99 (1996). https://link.aps.org/doi/10.1103/PhysRevLett.77.99